**Exploring EFL Students' Prompt Engineering in Human-AI Story Writing: An Activity Theory Perspective**

**Author information**

*Author #1 (corresponding author): David James Woo*

- Affiliation: Precious Blood Secondary School

- ORCiD: https://orcid.org/0000-0003-4417-3686

- Email: net_david@pbss.hk

*Author #2: Kai Guo*

- Affiliation: Faculty of Education, The University of Hong Kong

- ORCiD: https://orcid.org/0000-0001-9699-7527

- Email: kaiguo@connect.hku.hk

*Author #3: Hengky Susanto*

- Affiliation: Education University of Hong Kong

- Email: hsusanto13@gmail.com

**Disclosure statement**

The authors report there are no competing interests to declare.

**Data availability statement**

The data that support the findings of this study are openly available in Open Science Framework (OSF) at http://doi.org/10.17605/OSF.IO/J7RKY.

**Biographical note**

***David James Woo*** is a secondary school teacher. His research interests are in generative artificial intelligence and English language writing education.

***Kai Guo*** is a Ph.D. candidate in the Faculty of Education at The University of Hong Kong. His research focuses on second language writing, computer-supported collaborative learning, artificial intelligence in education, and gamification in education. His recent publications have appeared in international peer-reviewed journals such as *Computers & Education*, *Interactive Learning Environments*, *Journal of Educational Computing Research*, *TESOL Quarterly*, and *Assessing Writing*.

***Hengky Susanto*** received his BS, MS and PhD degree in computer science from the University of Massachusetts system. He was a postdoctoral research fellow at University of Massachusetts Lowell and Hong Kong University of Science and Technology. He was also senior researcher at Huawei Future Network Theory Lab. Currently, he is a principal researcher in a startup mode research laboratory and a lecturer at Education University of Hong Kong. His research interests include applied AI (computer vision and NLP), smart city, and computer networking.

**Exploring EFL Students' Prompt Engineering in Human-AI Story Writing: An Activity Theory Perspective**

**Abstract**

This study applies Activity Theory to investigate how English as a foreign language (EFL) students prompt generative artificial intelligence (AI) tools during short story writing. Sixty-seven Hong Kong secondary school students created their own generative-AI tools using open-source language models and wrote short stories with them. The study collected and analyzed the students' generative-AI tools, short stories, and written reflections on their conditions or purposes for prompting. The research identified three main themes regarding the purposes for which students prompt generative-AI tools during short story writing: a lack of awareness of purposes, overcoming writer's block, and developing, expanding, and improving the story. The study also identified common characteristics of students' activity systems, including the sophistication of their generative-AI tools, the quality of their stories, and their school's overall academic achievement level, for their prompting of generative-AI tools for the three purposes during short story writing. The study's findings suggest that teachers should be aware of students' purposes for prompting generative-AI tools to provide tailored instructions and scaffolded guidance. The findings may also help designers provide differentiated instructions for users at various levels of story development when using a generative-AI tool.

# 1. Introduction

Artificial intelligence (AI) natural language generation (NLG) tools such as ChatGPT have captivated popular imagination as they can produce impressive texts. The integration of NLG tools in education has generated many questions (Rospigliosi, 2023) and a growing interest among writing educators. This is because in educational contexts, many learners may struggle to write, not only because they may lack knowledge and skills to independently complete different text types (Latifi et al., 2021a), but also because they may be unable to independently face writing's cognitive demands (Latifi et al., 2021b). Besides, in large classroom contexts, students may not receive sufficient feedback from teachers or peers of equal status (Latifi et al., 2020). Thus, NLG tools may provide additional support to learners who struggle to write.

In the context of language learning, where learners may struggle to generate ideas and opinions (Hyland, 2019), struggle with grammar, vocabulary, syntax (De Wilde, 2023) and lack confidence in their writing abilities (Zotzmann & Sheldrake, 2021), NLG tools might act as a tutor for practicing language or an independent language learning medium (Haristiani, 2019), providing language learners with real-time feedback and support for various writing tasks (Chen et al., 2021). For instance, Guo et al. (2022) found students could interact with chatbots as a scaffold to better write argumentative essays. Additionally, researchers have found NLG tool-based activities can positively influence English as a foreign language (EFL) students' willingness to engage in English language (Tai & Chen, 2020). However, individual EFL students may perceive the affordances of using NLG tools differently and some may even perceive affordances as constraints (Jeon, 2022). For EFL students to effectively interact with NLG tools to complete English language writing tasks, it appears students will not only need strategies but also the right NLG tools (Woo et al., 2023).

Activity theory (AT; Engeström, 1987) provides a framework to analyze how EFL learners interact with NLG tools as a mediated activity system. The present qualitative study applies AT to explore the rules governing the use of NLG tools by EFL students to write short stories. By analyzing EFL students' written reflections for the rules they have developed to interact with NLG tools, the study can provide insights into human-AI collaboration in education, improving pedagogy and tool design.

## *1.1. Theoretical framework*

AT, rooted in the work of Vygotsky (1978) and Leontyev (1981), offers a useful framework for analyzing human activity in sociocultural contexts. Vygotsky (1978) posited that an individual is a *subject* that acts on an *object* by use of culturally-historically developed *tools.* This action leads to intended and unintended *outcomes*. Leontyev (1981) elaborated that such mediated activity is embedded in a social environment or *community* with an evolving *division of labor*. Engeström (1987) further conceptualized these elements of mediated activity and their structure of interactions. Importantly, he posited that human activity always occurs in a community comprising a division of labor and *rules* that govern how subjects use tools to achieve objects. Instructions, assumptions, and established practices are examples of rules.

Figure 1.

Activity system diagram

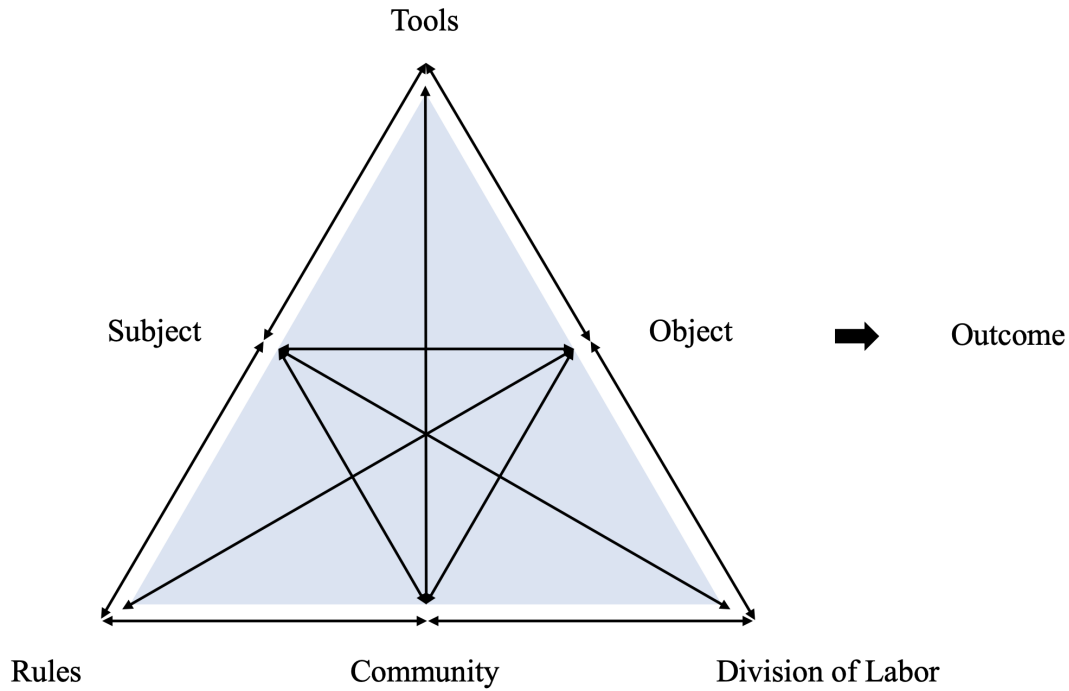

An essential mechanism to drive change and development of an activity system is *contradiction* (Engeström, 2001). According to Engeström (2001), contradiction refers to tension points within and between activity systems but not necessarily problems or conflicts. Engeström (1987) had defined four types of contradiction: a primary contradiction exists within an element in an activity system; a secondary contradiction exists between elements in an activity system, for instance, between a tool and an object; a tertiary contradiction exists between an old and a new activity system; and a quaternary contradiction exists between an activity system and neighboring systems. Attempts to resolve contradictions can lead to innovations and improvement of activity systems.

In the context of EFL story writing with NLG tools, AT assists researchers to conceptualize a mediated activity system: EFL students are subjects prosecuting their story writing. The stories are the objects and completed stories are an intended outcome. To complete their stories, students interact with their NLG tools. Within this perspective, although the subject, object and tool have basic relations, and exist in the same community, for instance, a school's classroom, we would expect different students individually completing their own stories to have different activity systems. This is because students would have different rules, or formal and informal conventions, by which they transform stories with NLG tools. Thus, each activity system would have its own division of labor, that is, responsibilities between student and tool.

Figure 2.

An individual student's activity system for completing a short story with NLG tools

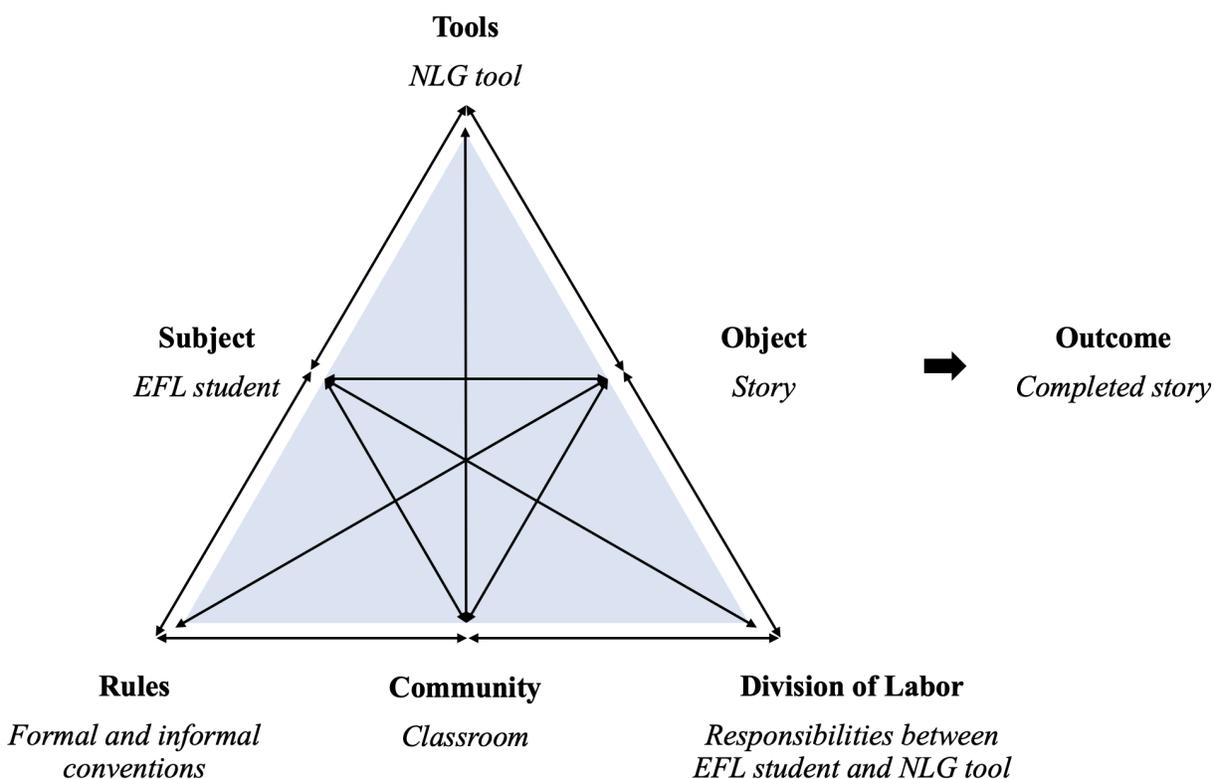

Within a student's activity system for completing a short story with NLG tools, we are interested in the interaction when, according to a student's writing objectives, the student writes a prompt to an NLG tool so that the tool generates text. Rules shape this human-AI interaction but at present we have little knowledge of these for EFL students. Since we have little specification of these rules, we define an EFL student's rules in terms of conditions or purposes for prompting the tool. Furthermore, it is important to note that various individual and contextual factors may influence EFL students' writing with NLG tools. For instance, Kuiken and Vedder (2021) have shown a correlation between students' written language proficiency and their academic achievement. Adams and Simmons (2019) conducted a study on gender differences in early writing and found that boys tended to produce shorter written compositions with fewer correctly spelled words, which were perceived as lower in quality compared to texts produced by girls. Guo et al. (2024) showed that students' writing with chatbots was influenced by specific writing conventions. Additionally, Fortunati and Vincent (2014) highlighted the effects of digital writing tools on students' writing outcomes. These studies indicate the need to understand how the different elements within the activity system of EFL students' writing with NLG tools interact with one another. By knowing EFL students' conditions or purposes for prompting the tool and the qualities of their other activity system elements, such as object, tool and community, we might compare neighboring activity systems that could inform the improvement of those systems.

### 1.2. Prompt engineering

A prompt to an NLG tool can be considered a set of instructions that programs the NLG tool to unlock its capabilities (White et al., 2023). Prompt engineering refers to the process of crafting a prompt that produces an NLG tool's most effective performance of a task (Liu et al., 2021). Liu et al. (2021) have defined two main varieties of prompts: a *cloze* prompt that leads a tool to fill in the blanks of a textual string; and a *prefix* prompt that leads a tool to extend a string prefix.

The content of a prompt greatly impacts an NLG tool's performance. For instance, changing the length of an input text, changing key words, or changing the order of words can impact the tool's performance. Although humans handcraft the most effective prompts (Zhou et al., 2023), there appears to be neither established practice nor much instruction for humans' prompt engineering: according to Dang et al. (2022), writing an effective prompt for an NLG tool is not straightforward for a non-technical user. They claim it has largely been a trial and error process for such a user. OpppenLaender et al. (2022) added that prompt engineering is a skill requiring expertise and practice to learn. Nonetheless, studies have reported some approaches to writing effective prompt content. For NLG tools (e.g., GPT-3) that have great capability to understand abstract task descriptions and human concepts, Reynolds and McDonell (2021) have proposed writing an explicit prompt in natural language with features such as direct task specification, demonstration, analogy and constraints. They also suggest these features can unlock an NLG tool's novel capabilities such as serializing reasoning, that is, breaking down a problem into steps before delivering a verdict, and meta-prompts, that is, writing task-specific prompts for itself. Furthermore, researchers have begun to demonstrate how prompt templates can assist people to realize their desired output from NLG tools (Strobelt et al., 2022).

Studies that have tested the generative capabilities of NLG tools for story writing have used story excerpts as prompt content. Clark and Smith (2021) had writers write a line of story alone as a prompt, after which the fusion- and GPT-2-based tool would generate the next line as an extension of that prompt. Similarly, Yang et al. (2022) input a writer's last written section of a story into a GPT-2-based tool, which would extend that section. Lee et al. (2022) had native English speaking adults write stories for which they selectively input the last story lines they had written as prompts for a GPT-3-based tool to generate an extension of those lines.

The above studies all adopt a turn-taking approach to human-AI interaction, where a human writer can voluntarily prompt an NLG tool and then use any of the tool's output in a story, but the studies have not provided any knowledge as to the conditions or purposes for a writer to prompt an NLG tool with a story excerpt. As far as we know, no study has been undertaken to identify the conditions or purposes for which an EFL writer would prompt an NLG tool to complete a short story writing task. Nonetheless, NLG tool designers have provided assumptions about human writers' possible purposes for which they would prompt a tool. For example, to test a LaMDA AI-powered story writing assistant with professional writers, Ippolito et al. (2022) gave writers the control over its assistant not only to extend a selected text, but also to elaborate a selected text, to suggest an alternative phrase for a selected text and to rewrite a selected text according to a specified property.

### 1.3. The study

As previously discussed, prior research has indicated the potential of students utilizing NLG tools to enhance their writing abilities. To effectively engage with these tools, students must develop various skills, including prompt engineering. Additionally, the use of NLG tools in the writing process can be influenced by a range of individual and contextual factors. However, our current understanding of students' activity system when using NLG tools for writing is still limited, as well as our knowledge of the key elements within the activity system that may impact their engagement in this activity. Therefore, the study aims to explore, first, the rules governing the use of NLG tools by EFL students in writing short stories, especially in view of the lack of established practices and instructions for this group of non-technical users' prompt engineering. Second, the study explores the qualities of the activity systems where these rules are active. The research questions (RQs) for this study are as follows:

RQ1: What are the conditions or purposes for which EFL students prompt an NLG tool during the task of story writing?

RQ2: What common qualities of activity systems do students with a common condition or purpose share, if any?

## 2. Methods

### 2.1. Research context and participants

We conducted the study during the 2022-23 academic year. Sixty-seven EFL Hong Kong secondary school students participated in the study. Fifty-seven of these students provided background information and their ages ranged from 12 ($n = 2$) to 17 ($n = 1$), with a median age of 14.4 and a mode of 14 ($n = 24$). Hong Kong secondary schools deliver curriculum to students in secondary grade levels one through six and the students' grade levels ranged from 1 ($n = 1$) to 5 ($n = 3$) with a mode of 3 ($n = 30$). When responding to the statement, "I have general knowledge about how AI is used today" on a six-point Likert scale, scores ranged from 1 (strongly disagree) ($n = 6$) to 6 (strongly agree) ($n = 1$) with a mode of 3 ($n = 20$) and mean of 3.1. On the same Likert scale, for the statement, "I think AI can help people write (for example, stories)," scores ranged from 1 ($n = 2$) to 6 ($n = 9$) with a mode of 4 ($n = 20$) and mean of 4.1.

The students came from four secondary schools in different Hong Kong geographic districts: 20 students came from a school that recruits students in the top third of academic achievement in its district; 31 students came from two schools that recruit students in the middle third of academic achievement in their districts; and 16 students came from a school that recruits students in the bottom third of academic achievement in its district.

The students attended two workshops designed to teach them to create NLG tools and use these tools to assist with their story writing. In the first workshop, students composed NLG tools using Python programming language and open-source AI on Hugging Face, a machine learning repository. Figure 3 shows the interface of a student's NLG tool where a student can input a prompt in the text box on the left side and the tool generates its output text on the right side. In the second workshop, students had 45 minutes to practice writing a story of no more than 500 words using their own words and words generated from their NLG tools. Students had the option to write a story from scratch or to rewrite an existing story. Figure 4 shows a story written using a student's own words in italics and red and words generated from the NLG tool from Figure 3 in non-italicized, black text.

Figure 3.

Interface of a student's NLG tool, called My First Text Generator, on Hugging Face (identifiers removed)

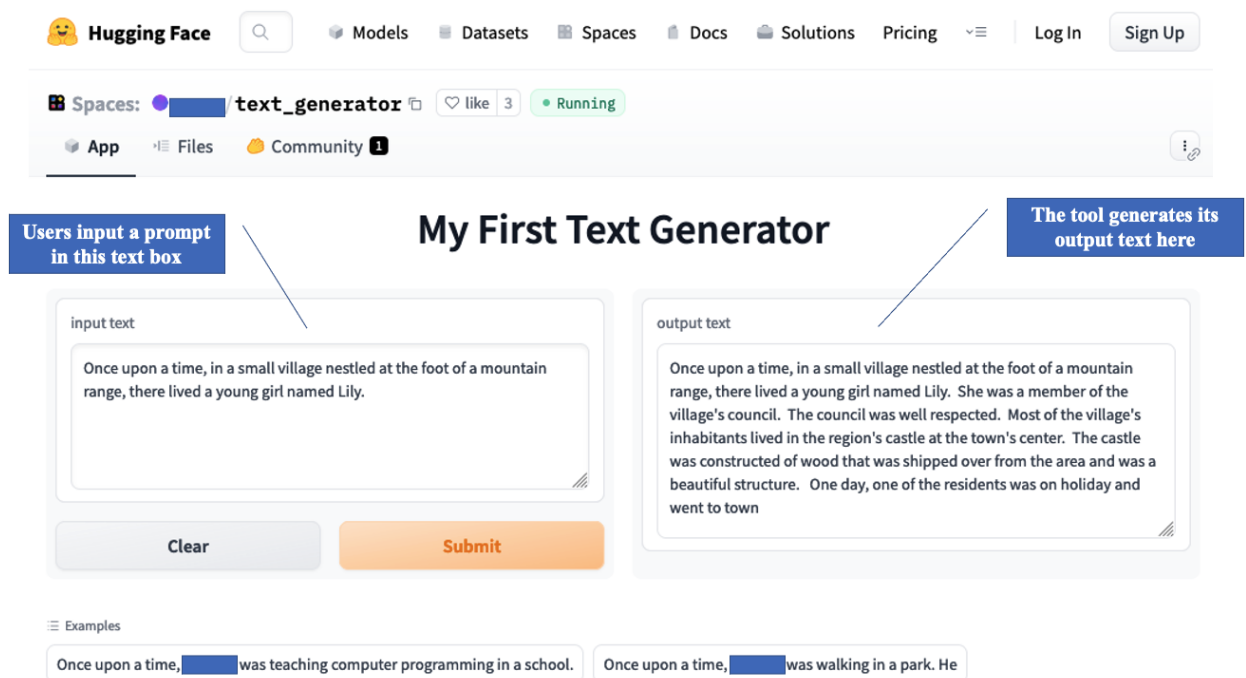

Figure 4.

A story written using a student's own words and words generated from their NLG tool

In the middle of an English lesson, a classmate who sat next to me, Tom, didn't feel well. *I asked him if he needed to drink some water. He said he did.* Late he said,," I'm okay, let me take a pill." The pill was red, with a heart on it. I asked him," What's that pill? There's a heart I'm on it, so cute!" He replied," It's just an antacid. I always take it when I'm not feeling well." Then, he put the pill in his mouth carefully.

After he had taken the pill, his limbs became twisted and his teeth became sharper, with his leg limping. "What's wrong with him? He looks so scary!" I shouted loudly.
Rushing to John, he bit him forcibly. We were all shocked and all shouted," There are zombies! Help!"
Bitten by Tom, John had blood all over his face. *His eyes were hollow and his hands were reaching to us.* Seeing Tom and John eager to bite us, we screamed. Our English teacher turned on her phone and dialed 999 for help. She told the police we were desperate for help. Hearing there were zombies, the police didn't believe us. We were frustrated.

At that moment, most of my classmates were bitten by them and became zombies. To let them return to normal, we tied them to the wall and gave them the pill that Tom had eaten. However, the pill didn't work. The survivors rushed out of the classmates and locked the door. Unfortunately, they escaped from the windows. *We ran along the corridor and yelled loudly.* Seeing zombies, students from other classes were astonished. The zombies rushed into other classes and bit them. After a while, the school was flooded with zombies. Chased by zombies, I ran into a room and locked the door. I saw my classmate, Sam, hiding under the table with his shaking hands. I took a picture of the school and posted it on social media. *The photo was spread widely.* Shocked by the photo, people commented that they had dialed 999 to help us.

15 minutes later, a helicopter landed on the rooftop. I asked Sam if he could run with me. He promised and he opened the door. *We looked around and ran along the corridor.* It was hard to run to the rooftop because the zombies were chasing after me. Therefore, I used some boxes to hit them when they wanted to bite me. Finally, we arrived at the rooftop. The rescuers comforted us, *I was touched that I survived.* We shouted elatedly," We're safe now!" When we were on the helicopter, it was shocking to see the playground turn into a zombie land. Also, I was sad that I couldn't study at that school anymore and almost all the students became zombies.

*Note.* The student's words are in italics and red; the words generated from their NLG tool are in non-italicized, black text.

### 2.2. Data collection

At the end of the second workshop, students completed an English language questionnaire on Google Forms. To collect data on the rules students have developed to interact with their NLG tools, in one open-ended question item we asked, "How did you decide which words, sentences or paragraphs to put into an NLG tool?"

To collect data on the common qualities of activity systems, we first selected activity system elements for which we could collect data and for which we could differentiate qualities. In this way, for the *tool* element, we asked students to share links to their NLG tools with us after the first workshop. For the *object* element, we asked students to share links to their stories with us after the second workshop. A total of 44 students shared their stories. For a *community* element, we had pre-existing data on each student's school and that school's level of academic achievement relative to other schools in the same school district.

Students were informed that their questionnaire responses, NLG tools and stories would be collected and analyzed for scientific purposes, their identities would be anonymized for study and they had the right to decline participation at any stage of study. Students provided their consent on the Google form questionnaire.

### 2.3. Data analysis

#### 2.3.1. For addressing RQ1

To address the first research question, we performed a thematic analysis (Braun & Clarke, 2006) on students' answers to the prompting question, "How did you decide which words, sentences or paragraphs to put into an NLG tool?". Our analysis aimed to uncover patterns within the students' answers, providing valuable insights into common themes in EFL students' conditions and purposes for prompting an NLG tool during story writing. To operationalize our analysis, we designed a coding scheme using an inductive approach, which enabled us to remain open to the data and to identify themes that were truly reflective of the students' answers, thereby enhancing the validity and relevance of our findings (Saldaña, 2012).

We aimed to capture students' considerations of what they wanted the NLG tool to do and what they wanted to achieve in prompting it. First, we read through all answers to get a sense of the range of ideas. We then conducted open coding, generating codes and assigning those codes to relevant excerpts. The coding process involved multiple iterations and two coders so that as we re-read the data and discussed, we refined, merged or separated codes. To enhance the trustworthiness and credibility of our final coding scheme, we created a codebook and conducted inter-coder reliability checks (see supplemental online material), comparing two coders' independent analysis and resolving any discrepancies through discussion and consensus (Miles et al., 2013). Table 1 lists the codes, their descriptions and examples, and inter-coder agreement information.

**Table 1**. Coding scheme

| (No.) Code | Label | Definition | Example excerpt | Non-example excerpt | Number of agreement instances | Number of text instances | Intercoder agreement |
|---|---|---|---|---|---|---|---|
| 1 | description | Any excerpt that contains the word **description**. | *Description about a certain object.* (student 1) | *Some adjective I would like to improve* (student 2) | 2 | 2 | 1.00 |
| 2 | don't know | Any excerpt that contains the expression, "I don't know" or intends to express "I don't know." This expression may indicate complete ignorance. | *Idk* (student 36) | *important words and sentences* (student 34) | 4 | 4 | 1.00 |
| 3 | elaboration | Any excerpt that contains the words, "**details**, "**elaborate**," "**expand**," or "**first**." | *when I want the description to be more detailed,* (student 11) | *by improvising* (student 10) | 8 | 8 | 1.00 |
| 4 | interesting | Any excerpt that contains the word "interesting." | *When i think it is useful or it can make my story interesting, i would put it into a text generator.* | *use conjunction* (student 49) | 2 | 2 | 1.00 |

(student 48)

| | | | | | | |
|---|---|---|---|---|---|---|
| 5 | more | Any excerpt that contains the word **more** or a similar word referring to a greater or additional quantity | *Some sentences that I want to write more and longer* (student 8) | *when I am not sure about how to start a new paragraph* (student 9) | 5 | 5 | 1.00 |
| 6 | my ideas | Any excerpt that contains the word **idea** and indicates that a student has possession of an idea, for instance, by the phrase "my idea" or "the idea." Also, "what // the things I want // think" The opposite of no idea code. | *Choose the word which is my original idea* (student 5) | *Copy from my writing plan* (student 6) | 9 | 10 | 0.90 |
| 7 | my story | Any excerpt that contains the phrase, "**my story**," "**that story**" or "**the story**," or a specific reference to a story feature, such as dialogs, scene and **setting** | *I like putting in some statements that introduce characters and situations so that I can see how the AI elaborates on my setting.* (student 20) | *When I had no ideas with what I am going to write I will put it into a text generator* (student 16) | 11 | 12 | 0.92 |

| 8 | new paragraph | Any excerpt that refers to starting a **new paragraph**. | *when I am not sure about how to start a new paragraph* (student 9) | *I put words that are suitable with my story.* (student 12) | 2 | 2 | 1.00 |
| 9 | no idea | Any excerpt that explicitly refers to **"no idea"** or a similar phrase, **"stuck on," "think about"** or **"think of"** | *Think about what I wanna write* (student 7) | *Some sentences that I want to write more and longer* (student 8) | 6 | 6 | 1.00 |
| 10 | part of speech | Any excerpt that refers to a part of speech such as an **adjective** or **conjunction** | *Some adjective I would like to improve* (student 2) | *Think about what I wanna write* (student 7) | 3 | 3 | 1.00 |
| 11 | plan | Any excerpt that refers to a **plan** or similar words, **"contents," "story line,"** and **"structure."** | *Copy from my writing plan* (student 6) | *Think about what I wanna write* (student 7) | 6 | 6 | 1.00 |
| 12 | words | Any excerpt that refers to **word** or **words** | *Think some words to fits the main idea for that story.* (student 29) | *I planned it* (student 25) | 7 | 7 | 1.00 |

After coding the data, we reread coded excerpts to help us gain a comprehensive understanding of the data and identify possible recurring patterns, themes, and ideas (Braun & Clarke, 2006). We looked for shared properties or relationships in the codes and grouped codes into broader categories or themes that captured the essence of the coded data. After refining and redefining the themes, we arrived at clear and concise descriptions of each theme's scope, focus, and significance. We ensured that they accurately represented the coded data and explored how the themes were connected to each other and to the research question (Braun & Clarke, 2006).

*2.3.2. For addressing RQ2*

To address the second research question, we operationalized variables within each selected activity system element to explore. First, we operationalized a *community* variable in terms of a student's school's overall level of academic achievement compared to other schools in the same district. We categorized that level of achievement as either *low*, *intermediate* or *high*.

We operationalized a *tool* variable in terms of the sophistication of a student's Python programming language used to create their tool. We categorized that level of sophistication as either *basic*, *intermediate*, or *advanced*. Basic refers to the elementary Python programming language taught in the workshop for building NLG tools. This includes important programming libraries and their dependencies, and using a single function. Intermediate refers to additional language taught in the workshop, including adding and changing parameters, instantiating variables and using multiple functions. Advanced refers to sophisticated language beyond what was taught including defining functions. For example, we categorized the Python language that composed the NLG tool in Figure 3 as advanced (see supplemental online material).

We operationalized *object* variables for a student's story in terms of the completeness of the story, the number of AI words and the overall quality of writing according to a standard scoring rubric. We analyzed stories only from 34 students who followed instructions to write no more than 500 words. We categorized each student's story as complete or incomplete. After measuring the number of AI words, we measured the overall quality of writing: two human experts independently scored each student's story for content, language and organization criteria according to a rubric (see Appendix 1) and without knowing which words came from the student and which came from their NLG tool; we averaged the content, language and organization scores from each expert and added those averaged scores to arrive at an overall quality of writing score for the student's story. For the number of AI words and overall quality of writing variables, we calculated a mean and a standard deviation. Then for each variable we categorized a student's story as either normal, that is, within one standard deviation; low or below one standard deviation; or high, that is, above one standard deviation. Using the story in Figure 4 as an example, we categorized the story as complete, the number of AI words as high and the overall quality of writing as normal.

From our cleaned activity system variable data (see Appendix 2), we created matrix displays (Miles et al., 2013). In each matrix display, the unit of analysis was a student, the *subject* in an activity system. First, we organized students according to the theme(s) attributable to their answers to the question item, "How did you decide which words, sentences or paragraphs to put into a text generator?" The themes are the row headings. Then we selected a variable of interest from another activity system element. The categories for that variable are the column headings. We analyzed the data by counting the number of students for each theme and variable category, creating bar charts, making comparisons and noting any salient patterns.

## 3. Findings

### *3.1. Themes for conditions or purposes (RQ1)*

From an AT perspective (Engeström, 1987), we aimed to explore rules for which EFL students prompted NLG tools during the task of story writing. After analyzing 67 students' answers, we found much variability in how much detail students provided in their answers. Some answers were just one-to-two words while others were full sentences. 15 students' answers did not seem to fit with any theme so we excluded these unclear or uninformative answers from the analysis.

Our thematic analysis revealed three main themes that can explain some conditions or purposes for EFL students' prompt engineering (Gatt & Krahmer, 2018). The three themes provide the best synthesis of the patterns found in students' answers. Since the themes represent an integration of patterns across many student answers, the themes are not mutually exclusive. Some students' answers reflect multiple themes. In the following sections, we elaborate each theme. We report the theme's name, the code(s) from which it was composed, the number of students showing the theme in their answers, a description of the theme and representative quotes of each theme.

### 3.1.1. Theme 1: Unawareness of rules

The first theme shows four students prompted NLG tools without a clear understanding of when or why they should do so. For the reflection question, these students answered, "I don't know" (student 21), "idk" (student 36) and "I done not" (student 37).

### 3.1.2. Theme 2: Overcoming writer's block

Eight students' common purpose for prompting was overcoming writer's block or generating new ideas when these students felt stuck. Within this theme, students showed metacognitive awareness of their lack of ideas, being unable to move forward with their story. They provided answers like, "When I am stuck on constructing a complete scene" (student 19), and "When I had no ideas with what I am going to write I will put it into a text generator" (student 16). Furthermore, some students showed awareness of precisely where in their story they needed new ideas. They provided answers like, "when I am not sure about how to start a new paragraph" (student 9) and, "I decide to choose the word with good meaning to open another paragraph" (student 46).

### 3.1.3. Theme 3: Developing, expanding and improving the story

We found 44 students' common purpose for prompting was to enhance, expand, and further develop their stories. We identified nuance within this theme.

*Further development of pre-existing ideas or plan*

We found 34 students had pre-existing ideas or plans for their stories and used NLG tools to further develop them. They provided answers like, "I prefer putting in more complete sentences so that the AI gives more related paragraphs I can use in my story" (student 18) and, "Think some words to fit the main idea for that story" (student 29). Of these students, some appeared to have more structured ideas and referred to plans. They provided answers like, "Copy from my writing plan" (student 6) and, "By following the story line" (student 60).

On the other hand, it appeared some students had less structured ideas. They referred only to words, providing answers like, "Write the sentences or words on the tool" (student 42) and, "I put words that are suitable with my story" (student 12).

*Improving story with more detailed description and better elaboration*

We found 17 students' purpose for prompting was improving their story through detail and elaboration. Students provided answers like, "when I want the description to be more detailed" (student 11) and, "That I want to elaborate more" (student 55). Of these students some appeared to be looking for a particular descriptive word. For instance, one student answered, "Some adjective I would like to improve" (student 2) and another, "adjective" (student 64).

On the other hand, some students sought to enhance particular story aspects, providing answers like, "I like putting in some statements that introduce characters and situations so that I can see how the AI elaborates on my setting" (student 20). Other students intended to make their story more interesting through additional details. These students provided answers like, "When I think it is useful or it can make my story interesting, I would put it into a text generator" (student 48) and, "When I wanted more ideas and more interesting sentences I would use the AI to help" (student 32).

### 3.2. Common qualities of activity systems (RQ2)

Fifty-two students provided answers that were coded to at least one theme. Of these students, two students' answers were coded to the themes of overcoming writer's block and developing, expanding and improving the story. Thus, within the *rule* element, few students showed more than one condition or purpose for prompting an NLG tool.

When we analyzed the data of students for each theme with variable data of other activity system elements, we observed some patterns that may indicate common qualities of activity systems for students of the same theme. When comparing patterns across themes, we observed possible differences in common activity system qualities between students of different themes. We present our analysis as bar charts with prose. We note 20 students had incomplete data for at least one activity system element (see Appendix 2) and any incomplete data was not included in any bar chart analysis.

For the *community* element (see Figure 5), we observed the majority of *unawareness* theme students come from low academic achievement schools and none come from high academic achievement schools. In contrast, students with *overcoming* or *developing* themes were found in schools at all levels of academic achievement.

Figure 5.

*Community* element

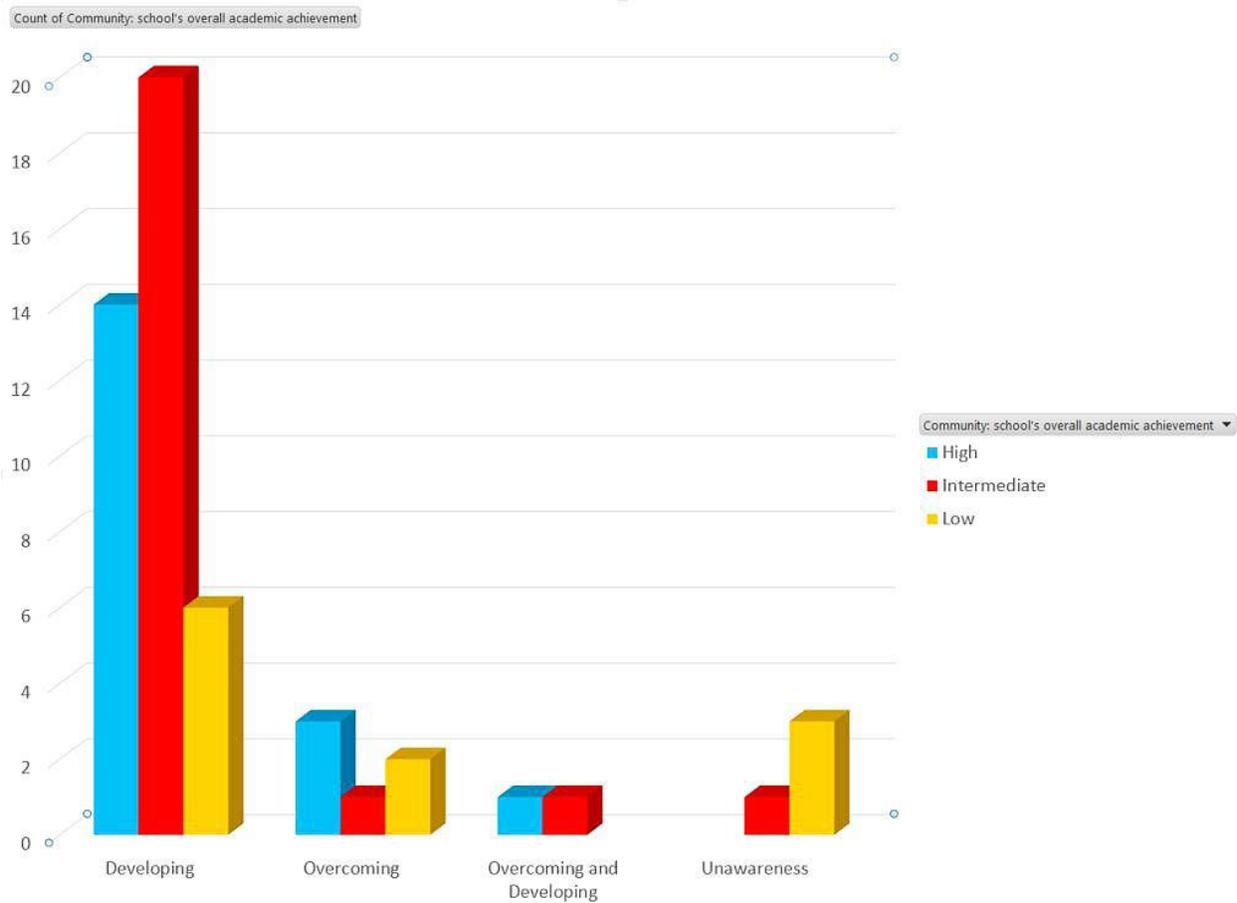

For the *tool* element (see Figure 6), we observed students with the *developing* or *unawareness* theme used tools at all levels of Python programming language sophistication. In contrast, students with the *overcoming* theme used tools only at the intermediate and advanced levels.

Figure 6.

*Tool* element

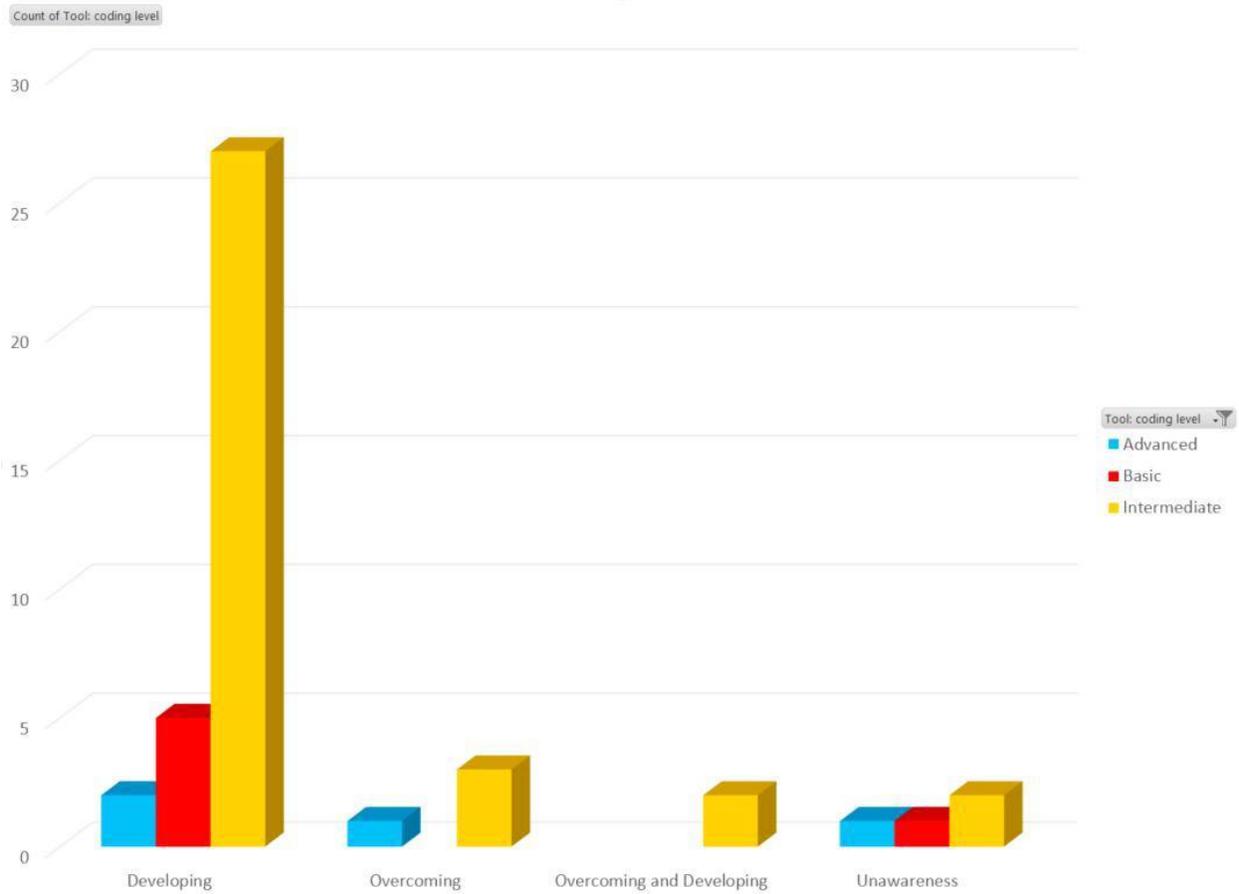

For the *object* element's completeness of story variable (see Figure 7), we observed the vast majority of *overcoming* and *developing* theme students shared complete stories with us. In contrast, half of the *unawareness* theme students shared incomplete stories with us.

Figure 7.

*Object* element's completeness of story variable

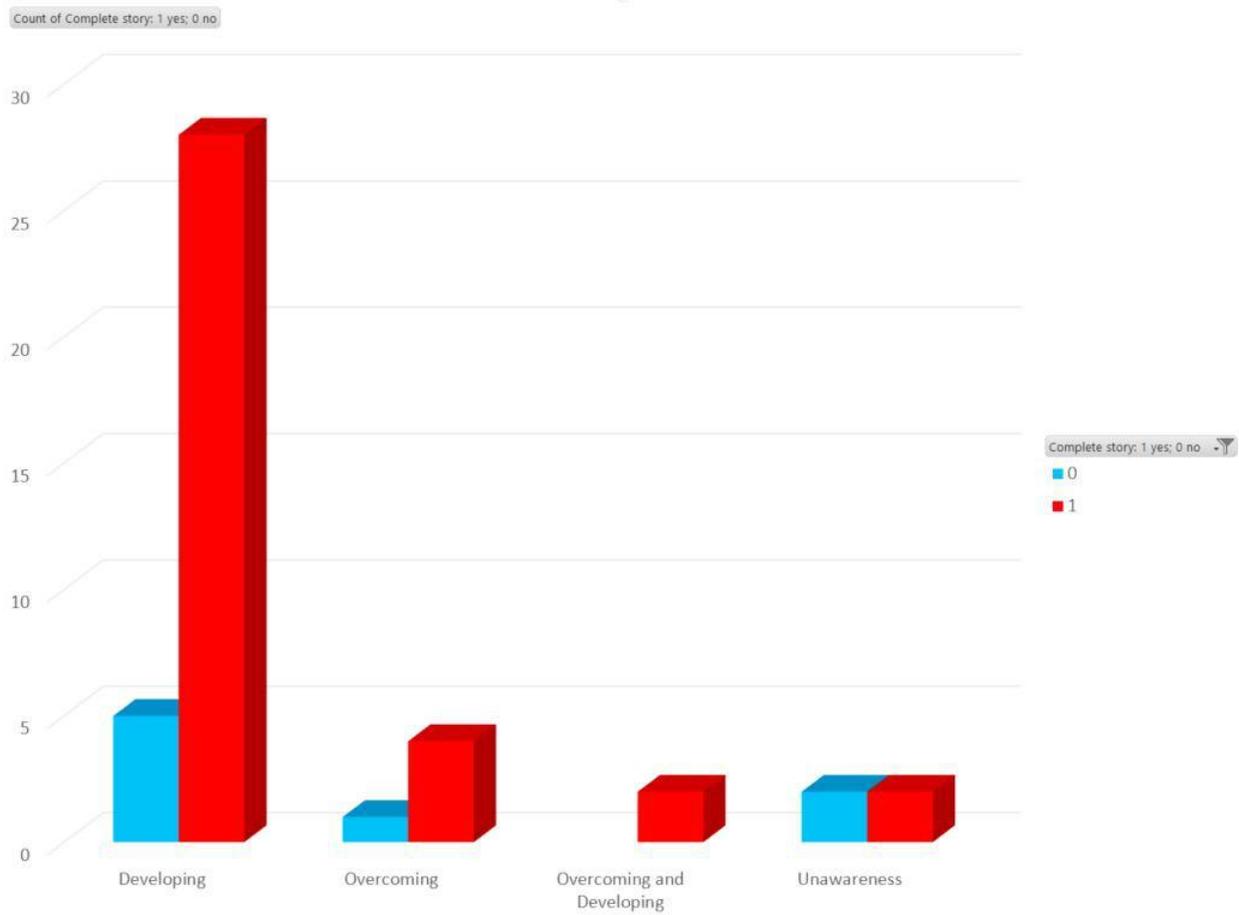

For the variable number of AI words (see Figure 8), we observed *unawareness* theme students used AI words at low and normal levels, *overcoming* theme students at normal and high levels and *developing* theme students at all levels.

Figure 8.

Variable number of AI words

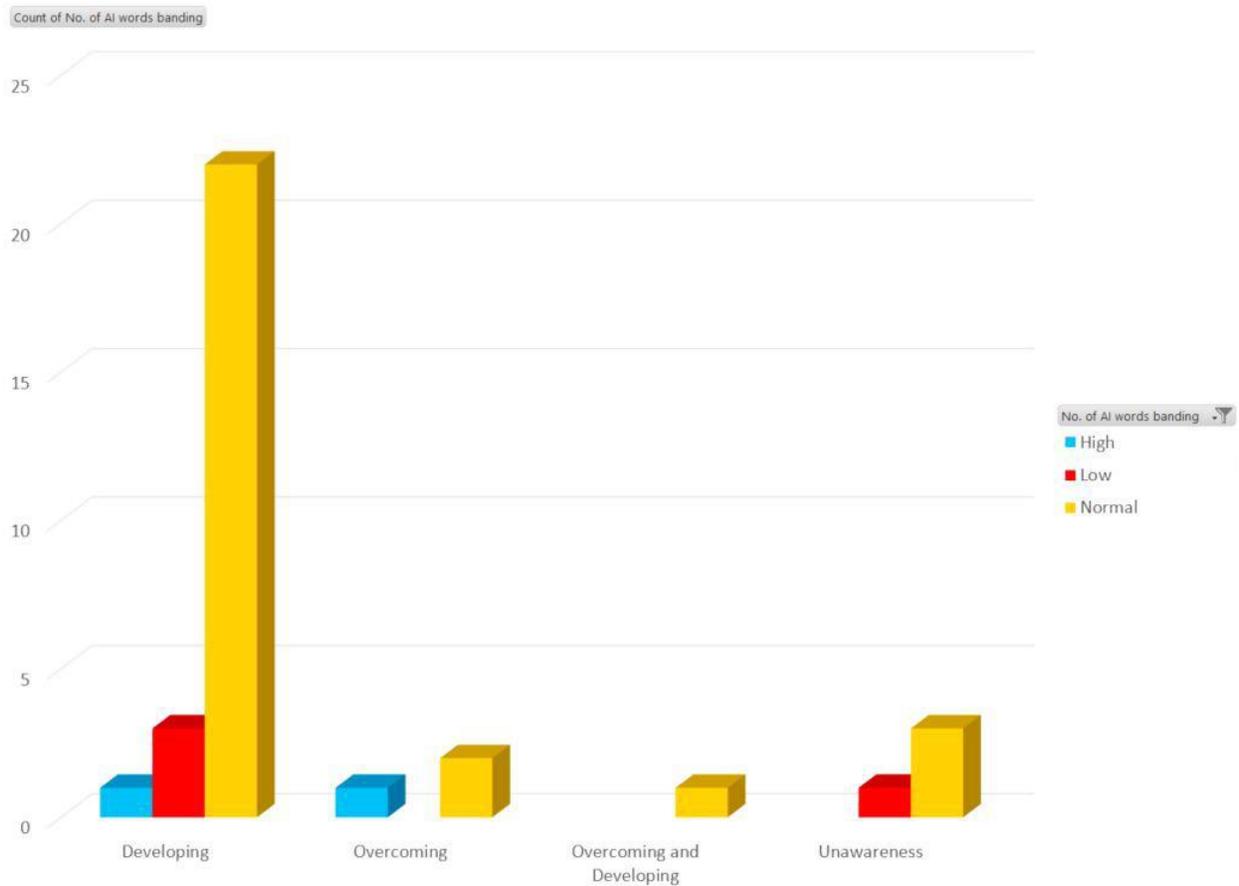

For the variable overall writing quality scored according to the standard rubric (see Figure 9), we observed the majority of *unawareness* theme students scored low and none scored high. In contrast, *overcoming* theme students scored normal or high, none low; and *developing* theme students were found at all levels of overall writing quality.

Figure 9.

Overall writing quality scored

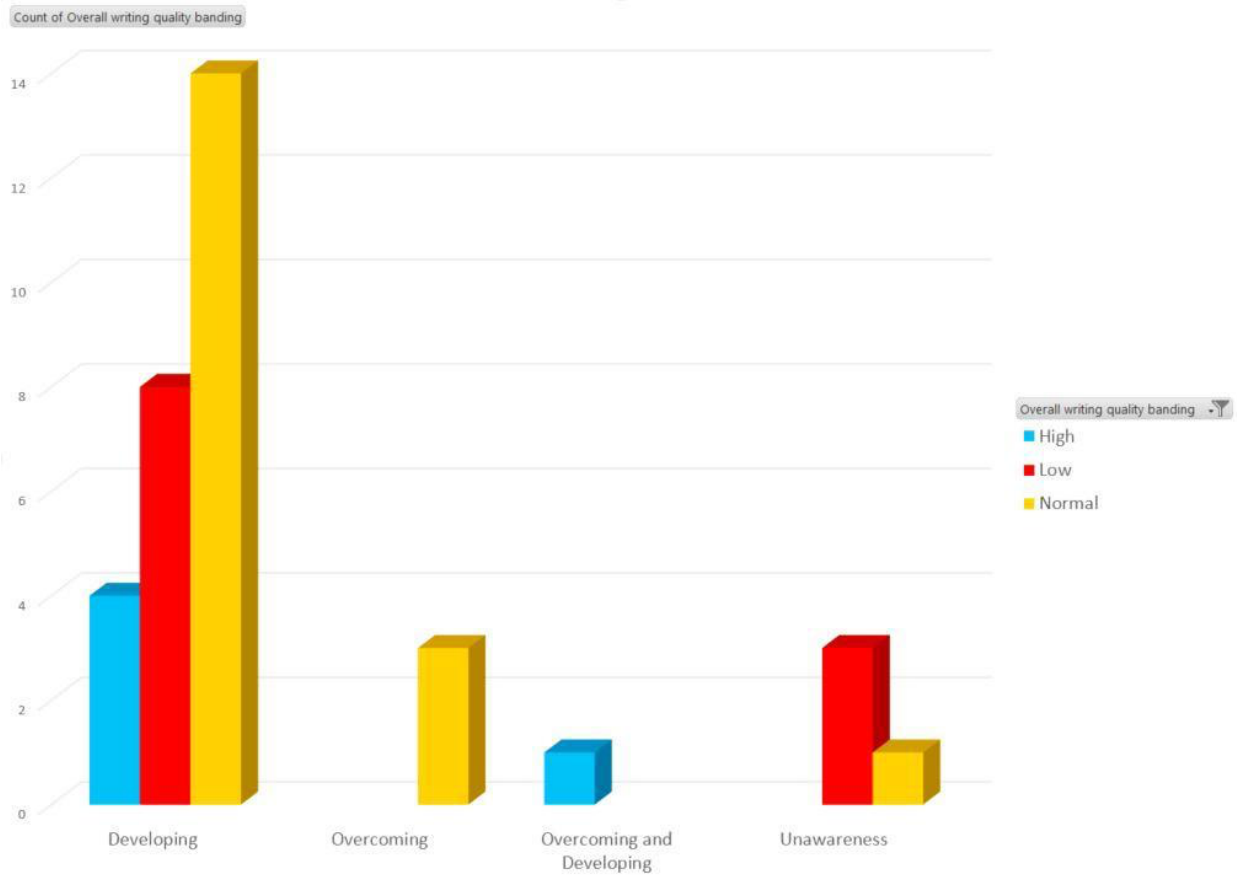

## 4. Discussion

The present study explored EFL students' prompt engineering in human-AI story writing using

AT (Engeström, 1987) as a theoretical framework. From an AT perspective, the qualities of an

activity system shape how subjects work to achieve the object. For the first research question, the

findings revealed three main themes on conditions or purposes for which EFL students prompt

NLG tools during the task of short story writing: unawareness of rules; overcoming writer's block; and developing, expanding, and improving the story. For the second research question, the findings provide insights into the common qualities of activity systems for EFL students prompting AI tools for specific conditions or purposes during story writing. The common qualities we identified for students of each theme suggest the kinds of activity systems in which those students were operating.

The first theme showed a few students lacked explicit knowledge about when and why to prompt NLG tools. This theme shows that students may not always understand the use of AI tools in educational contexts. Like requiriing students to engage in peer learning processes without appropriate support (e.g. checklists; prompts; and tutorials) (Latifi et al., 2021b), requiring students to use NLG tools without such support may fail to realize the tools' potential in these struggling students' writing. Besides, these EFL students may also be severely limited users of English language, another essential tool in an activity system (Vygotsky, 1978). Within this cohort, some students' common qualities of low academic achievement schools, use of basic tools, and low overall writing quality suggest they worked within precarious or poorly-developed activity systems. Some students' incomplete stories and low use of AI words indicate they may struggle to achieve the object, perhaps due to a lack of AI and English language knowledge and skills. Within a precarious activity system, our study highlights a tension of EFL students likely lacking technological and language exposure in a low-performing community. It also highlights the importance of teachers and other communities to provide appropriate support in terms of language instruction and NLG tool instruction to assist these students in completing writing tasks. If not, students may develop an over-reliance on unethical use cases of NLG tools, such as using NLG text to replace human effort in writing, not least threatening academic integrity and inhibiting the development of students' own higher order cognitive skills (Farrokhnia et al., 2023).

The second theme of "overcoming writer's block" showed that some students prompted NLG tools for idea generation, overcoming obstacles to complete their story writing. This finding aligns with previous research showing that NLG tools can facilitate writer's idea generation (Clark & Smith, 2021; Yang et al., 2022). NLG tools may resemble other online learning environments where learners can receive constructive ideas, thereby reducing pressure on actual peers and teachers for feedback in the writing process (Latifi et al., 2020). Compared to students with the first theme, more students with the second theme worked with more advanced tools and scored higher in overall writing quality. In contrast to the students with the first theme, these students with the second theme may have worked within robust activity systems that better supported achieving the object. Besides, a key difference between students with themes 1 and 2 is the latter's metacognitive awareness to mediate their activity with tools to achieve objects.

The third theme of "developing, expanding, and improving the story" showed many students had pre-existing ideas or plans. Compared to other students in the study, these students most strategically used NLG tools to advance their goals. Students with this theme most clearly highlight the potential of human-AI story writing to refine and enhance story development (Lee et al., 2022). These students' use of NLG tools might transform the quality of their stories, which highlights the importance of integrating NLG tools in EFL classroom writing so as to facilitate students' achievement of higher levels of creativity and English language proficiency in writing tasks (Dai, 2010). Nonetheless, although these EFL students appear to have the most sophisticated conditions or purpose of students in this study, their rule appears far less sophisticated than those suggested by Ippolito et al. (2022) for professional writers. This suggests that even the most advanced writers in an EFL classroom context can benefit from instruction to refine their conditions and purposes for prompting to a finer-grain level and with higher-order thinking. In other words, these writers may benefit from additional prompting directions and strategies, or feedforward feedback (Latifi et al, 2021b), from peers and teachers. Moreover, the students with theme 3 spanned all levels for most activity system qualities, suggesting a diversity of activity systems. This aligns with Engeström's (2001) notion of activity systems in constant flux and evolution.

## 5. Theoretical contributions and pedagogical implications

Our study's noteworthy theoretical contributions are, first, its showcasing the practical application of the AT framework in the specific research context of studying EFL students' writing with NLG tools. By employing AT, we provide a valuable contextualization of this theoretical framework. Our findings can spur future research to explore the application of AT in diverse AI in education contexts. By employing AT in these different learning contexts, researchers can gain insights into students' utilization of AI and its impact on learning outcomes. Secondly, our findings shed light on the intricate activity system of EFL students' writing when utilizing NLG tools. This provides valuable insights into the interplay and dynamics among different elements within the activity system. Future studies could delve deeper into exploring the relationships among these elements and examine how conflicts within the system arise and influence its development. Such investigations would contribute to a more comprehensive understanding of the activity system and its evolution in the context of EFL students' writing with NLG tools.

Additionally, the findings of this study on EFL students' prompt engineering rules for human-AI story writing can provide the following implications for pedagogy. First, the findings may help teachers to better understand the emergent purposes and conditions that shape students' collaboration with NLG tools. Teachers can gain insights into individual students' tendencies, difficulties and strategies in prompting NLG tools. Furthermore, the findings on common qualities of activity systems for different themes can enable teachers to better situate or contextualize individual student's prompt engineering rules. With such knowledge, teachers can provide tailored instructions and guidance to facilitate students' effective prompt engineering. For instance, teachers may provide more guidance if students have no idea when and how to prompt AI, but less guidance if students can articulate clear writing plans and ideas for AI's further elaboration. Ultimately, teachers are in a position to introduce the tension in a student's activity system that leads to system improvement and language learning.

Insights into students' conditions and purposes for prompting NLG tools for short story writing can enable designers of intelligent tutoring systems (Chen et al., 2021) to deliver differentiated instructions for different user levels. Besides, NLG tools that provide more structure assistance (Gayed et al., 2022) are especially helpful in classroom writing contexts when teachers have limited knowledge of NLG tools or when they face constraints in providing tailored instruction and guidance to individual students. Based on our study, a tool can present instructions to students on when or for what reasons students might prompt the tool. Second, a tool can present suggestions for prompt content based on specific conditions or purposes, for instance, when students have no idea about what to write or when students want elaboration of specific story content. Importantly, for EFL learners and for more capable NLG tools, instructions for students can be presented in different languages, and prompt content can be received in different languages.

## 6. Limitations and future directions

In summary, our exploratory study provides a set of themes that express rules for EFL students to prompt generative-AI for story writing. Our study's methodological limitations can facilitate further research. First, as our sample size was small (N = 67), with a larger dataset of student answers, we may identify additional themes or subthemes. Thus, this study's themes are not exhaustive of all conditions or purposes for which students prompt generative-AI for story writing. Second, since we present themes representative of students' thoughts from a single self-reflection prompt, by expanding the context, with more questions or other data (e.g., observations and interviews), we could gain a richer understanding of students' conditions and purposes for prompting generative-AI for story writing. Likewise, since we had delivered the prompting question in the English language we could gain a richer understanding by delivering the prompting question in students' native languages and receiving more elaborate answers in their native languages. Third, our analysis looked at students as a whole group but did not consider how conditions or purposes may differ based on factors like gender, English writing proficiency, digital literacy, or teaching quality. These individual differences could influence the themes. Similarly, further research can evaluate whether some reported conditions or purposes for prompting generative-AI are associated with higher quality short stories.

**Appendix 1**. Assessment rubric

| Score | Content | Language | Organization | AI Words |
|---|---|---|---|---|
| 5 | · Content fulfills the requirements of the question<br>· Almost totally relevant<br>· Most ideas are well developed/supported<br>· Creativity and imagination are shown when appropriate<br>· Shows general awareness of audience | · Wide range of accurate sentence structures with a good grasp of simple and complex sentences<br>· Grammar mainly accurate with occasional common errors that do not affect overall clarity<br>· Vocabulary is wide, with many examples of more sophisticated lexis<br>· Spelling and punctuation are mostly correct<br>· Register, tone and style are appropriate to the genre and text-type | · Text is organized effectively, with logical development of ideas<br>· Cohesion in most parts of the text is clear<br>· Strong cohesive ties throughout the text<br>· Overall structure is coherent, sophisticated and appropriate to the genre and text-type | · AI words compose less than ⅓ of the total number of words in the text |
| 4 | Between 3 and 5 | | | |
| 3 | · Content just satisfies the requirements of the question<br>· Relevant ideas but may show some gaps or redundant information<br>· Some ideas but not well developed<br>· Some evidence of creativity and | · Simple sentences are generally accurately constructed.<br>· Occasional attempts are made to use more complex sentences. Structures used tend to be repetitive in nature<br>· Grammatical errors sometimes affect meaning<br>· Common vocabulary is generally appropriate<br>· Most common words are spelt correctly, with basic | · Parts of the text have clearly defined topics<br>· Cohesion in some parts of the text is clear<br>· Some cohesive ties in some parts of the text | · At least 8 AI chunks of any length |

| | | | |
|---|---|---|---|
| | imagination<br>· Shows occasional awareness of audience | punctuation being accurate<br>· There is some evidence of register, tone and style appropriate to the genre and text-type | · Overall structure is mostly coherent and appropriate to the genre and text-type | |
| 2 | Between 1 and 3 | | | |
| 1 | · Content shows very limited attempts to fulfill the requirements of the question<br>· Intermittently relevant; ideas may be repetitive<br>· Some ideas but few are developed<br>· Ideas may include misconception of the task or some inaccurate information<br>· Very limited awareness of audience | · Some short simple sentences accurately structured<br>· Grammatical errors frequently obscure meaning<br>· Very simple vocabulary of limited range often based on the prompt(s)<br>· A few words are spelt correctly with basic punctuation being occasionally accurate | · Parts of the text reflect some attempts to organize topics<br>· Some use of cohesive devices to link ideas | · AI words used in long chunks (more than 1 sentence in length) and in short chunks (less than 5 words in length). |

*Note.*

1. The rubric descriptors for content, language and organization levels 5, 3 and 1 are taken from HKDSE English writing rubric descriptors for levels 6, 4 and 2, respectively.

2. Content mark cannot exceed 1 if the story is not complete, that is, missing exposition; conflict; climax; and / or resolution.

3. Content mark cannot exceed 1 if the story is not a story, for example, an article or an essay

4. Creativity in content refers to the details, transformation and originality of ideas

5.  Language and organization marks cannot exceed +/- 1 of the content mark.

**Appendix 2**. Cleaned activity system variable data

| Student No. | Rule: theme | Community: school's overall academic achievement | Tool: coding level | Object: | | |
|---|---|---|---|---|---|---|
| | | | | Complete story: 1 yes; 0 no | No. of AI words banding | Overall writing quality banding |
| 1 | Developing | High | Intermediate | 1 | Normal | Normal |
| 2 | Developing | High | Intermediate | 1 | / | / |
| 3 | Developing | High | Intermediate | 1 | Normal | Normal |
| 4 | Developing | High | Intermediate | 0 | Normal | Low |
| 5 | Developing | High | Intermediate | 1 | Normal | Normal |
| 6 | Developing | High | Advanced | 1 | Normal | Normal |
| 7 | Overcoming | High | Advanced | 1 | High | Normal |
| 8 | Developing | High | Advanced | 1 | Normal | Normal |
| 9 | Overcoming | High | Intermediate | 1 | / | / |
| 11 | Developing | High | Intermediate | / | / | / |
| 12 | Developing | High | Intermediate | 1 | / | / |
| 13 | Developing | High | Intermediate | 1 | Normal | High |
| 14 | Developing | High | Intermediate | 1 | Normal | Normal |
| 16 | Overcoming | High | Intermediate | 1 | Normal | Normal |
| 17 | Developing | High | / | 0 | Low | Low |
| 18 | Developing | High | Intermediate | 1 | Normal | Normal |
| 19 | Overcoming and Developing | High | Intermediate | 1 | Normal | High |

| 20 | Developing | High | Intermediate | 1 | Normal | High |
|---|---|---|---|---|---|---|
| 21 | Unawareness | Intermediate | Intermediate | 0 | Low | Low |
| 23 | Developing | Intermediate | Intermediate | 1 | / | / |
| 24 | Developing | Intermediate | Basic | 1 | Normal | Normal |
| 25 | Developing | Intermediate | Intermediate | 1 | / | / |
| 26 | Developing | Intermediate | Basic | 0 | Normal | Low |
| 27 | Developing | Intermediate | Basic | 0 | Normal | Low |
| 28 | Developing | Intermediate | Basic | 1 | Normal | Low |
| 29 | Developing | Intermediate | / | / | / | / |
| 31 | Developing | Intermediate | / | / | / | / |
| 32 | Developing | Intermediate | / | / | / | / |
| 34 | Developing | Low | Intermediate | 1 | Normal | Low |
| 36 | Unawareness | Low | Basic | 0 | Normal | Low |
| 37 | Unawareness | Low | Advanced | 1 | Normal | Normal |
| 41 | Developing | Low | Intermediate | / | / | / |
| 42 | Developing | Low | / | 1 | / | / |
| 44 | Developing | Low | Basic | 1 | Normal | Normal |
| 45 | Overcoming | Low | / | / | / | / |
| 46 | Overcoming | Low | / | 0 | / | / |
| 47 | Unawareness | Low | Intermediate | 1 | Normal | Low |
| 48 | Developing | Low | Intermediate | 1 | Normal | Normal |
| 49 | Developing | Low | Intermediate | / | / | / |
| 51 | Developing | Intermediate | Intermediate | / | / | / |

| 52 | Developing | Intermediate | Intermediate | 0 | Normal | Low |
|----|-----------|--------------|--------------|---|--------|-----|
| 53 | Developing | Intermediate | / | 1 | Normal | High |
| 55 | Developing | Intermediate | Intermediate | 1 | High | Low |
| 56 | Developing | Intermediate | Intermediate | 1 | Normal | High |
| 57 | Overcoming | Intermediate | Intermediate | 1 | Normal | Normal |
| 58 | Developing | Intermediate | Intermediate | 1 | / | / |
| 60 | Developing | Intermediate | Intermediate | 1 | Low | Normal |
| 61 | Developing | Intermediate | Intermediate | 1 | Normal | Normal |
| 62 | Overcoming and Developing | Intermediate | Intermediate | 1 | / | / |
| 64 | Developing | Intermediate | Intermediate | 1 | Normal | Normal |
| 66 | Developing | Intermediate | Intermediate | 1 | Low | Normal |
| 67 | Developing | Intermediate | Intermediate | 1 | / | / |